# Wikipedia traffic data and electoral prediction: towards theoretically informed models

Taha Yasseri[1,2] and Jonathan Bright[2]

Oxford Internet Institute, University of Oxford, 1 St Giles' Oxford OX13JS, UK

taha.yasseri@oii.ox.ac.uk and jonathan.bright@oii.ox.ac.uk

**Abstract**

This aim of this article is to explore the potential use of Wikipedia page view data for predicting electoral results. Responding to previous critiques of work using socially generated data to predict elections, which have argued that these predictions take place without any understanding of the mechanism which enables them, we first develop a theoretical model which highlights why people might seek information online at election time, and how this activity might relate to overall electoral outcomes, focussing especially on how different types of parties such as new and established parties might generate different information seeking patterns. We test this model on a novel dataset drawn from a variety of countries in the 2009 and 2014 European Parliament elections. We show that while Wikipedia offers little insight into absolute vote outcomes, it offers a good information about changes in both overall turnout at elections and in vote share for particular parties. These results are used to enhance existing theories about the drivers of aggregate patterns in online information seeking.

## 1. Introduction

As digital technologies become more and more integrated into the fabric of social life their ability to generate large amounts of information about the opinions and activities of the population increases. The potential applications of what we will call here "socially generated data", which range from mobile phone GPS records to content produced on Twitter and searches made on Google, are widespread in terms of new understanding of human behaviour [1,2]. One of the liveliest areas of research which has been generated is on the possibility of using these data for predicting a wide range of social phenomena [3] such as stock market fluctuations [4,5], outcomes in public health [6,7], movie box office revenues [8], unemployment [9], and election results [10–14]. Social media data have also been used to gauge and quantify the popularity [15] and attention given to political parties [16]. The potential opportunities in this area are enormous: predictions based on socially generated data are much cheaper than conventional opinion polling, offer the potential to avoid classic biases inherent in asking people to report their opinions and behaviour, can deliver results much quicker and be updated more rapidly, and offer potential purchase on phenomena (such as stock market movements) for which there are no well-established forecasting models.

However, whilst a variety of positive results have been published, some strong criticisms of social generated predictions have also been raised, particularly in the field of electoral prediction (which, perhaps because of the widespread availability of validation data or because of its importance for conventional opinion polling, has been the most frequent application of this type of prediction). Most importantly, several authors have highlighted that work which has shown correlations between socially generated data and electoral outcomes has been done "without an analysis of what principle enables them" [17]. For example, given what is known about how Twitter's user base differs in demographic characteristics from the general population [18,19], and also the fact that people using Twitter do not necessarily have an incentive to tweet about their voting intention, there is little theoretical reason to expect that the overall volume of tweets about different politicians will be proportionate to their overall volume of votes. This has led several authors to argue that the positive correlations which have been observed so far between socially generated data and opinion polling could have easily been achieved by chance [20], with Metaxas et al. arguing that "predicting elections with accuracy should not be without some clear understanding of why it works" [21].

---
[1] Corresponding Author
[2] Equal Contribution

Our approach in this article is inspired by the challenge laid down by some of these critiques. The aim is to develop a theoretically informed prediction of election results from socially generated data, which is based not just on observation of correlation between raw numbers and eventual outcomes but also an understanding of the social processes through which the numbers are generated. We also aim to apply the model to a variety of different countries in two separate elections, thus offering a more general test of their usefulness than work which has focussed on just one country. Through this process we hope to both explore the predictive power of socially generated data and enhance confidence that the reasons for these predictions are knowable.

We focus on the readership statistics of politically relevant Wikipedia articles (such as those of individual political parties) in the time period just before an election. We will hence begin by outlining a theory of the relationship between Wikipedia page view statistics and overall electoral outcomes. We base this theory on a rational choice approach to explaining voting behaviour [24], which conceptualises voters as similar to consumers in a market, seeking to vote for the political party who offers them the greatest "pay-off" in terms of policies[3]. Online information seeking, from this rational choice perspective, can be explained in terms of voters looking for more information about the election: perhaps about practical matters such as how to vote, or perhaps about substantive matters such as which political party might suit them best. Such information seeking is rational in that it increases the chance that the voter will vote for the party which represents them best.

If this theory is correct, increases in Wikipedia page views ought to indicate increases in turnout at election time (both in terms of overall turnout of voters at the election and turnout for specific parties), as they should indicate that more voters are considering voting in the election for a given party. This logic is analogous to the logic behind the use of Wikipedia page views to forecast consumer behaviour in other areas such as stock market movements [4] and movie box office revenues [8]: while we do not know if any individual reader will actually decide to vote for a given party, a greater pool of people reading about the party ought to indicate a greater pool of potential voters, which should indicate more votes. Of course, Wikipedia page view traffic will also include journalists writing stories about the parties, "opinion leaders" looking for information to pass on to their social connections [citation], and those working for the parties themselves checking to see pages are up to date, however we assume the number of these people is small in relative terms, and hence do not include them in the theory

However, we also expect that a number of factors will moderate the relationship between online information seeking and electoral outcomes, meaning that Wikipedia page views may not be much use as a direct predictor. Rational voters are "cognitive misers" [25] who will not seek new information unless they think it necessary, thus minimizing the costs of voting. This has a number of potential consequences. People are more likely to seek information on new political parties, as they are less likely to have a pre-established opinion on whether this party offers them a good pay-off. However, they are also likely to consider the perceived "viability" of a party (i.e. whether it is likely to be able to contend as a serious electoral force): people are less likely to seek information on new parties which do not appear to have a chance of seriously contending [26] (this may also serve to create a kind of "spiral of silence" around minor parties who, because they are perceived to be less viable, are systematically ignored when people seek information [27]).

---

[3] Other approaches such as socio-demographic and psychological approaches to voting offer less clear expectations about why people should seek information online.

In addition to being more likely to seek information on new parties, people are also more likely to seek information if they are considering changing their vote. This might emerge from dissatisfaction with the party for whom they voted at the previous election (for example, studies have shown that dissatisfaction with government can promote higher online information seeking [28]). Hence Wikipedia page views may be driven more by "swing voters" who are switching to a new party, and who wish to inform themselves about their choices, rather than voters who are voting for the same one again. Recent studies have shown support for the idea that swing voting is associated with information seeking by showing that these voters typically have at least some amount of political knowledge [29], and that very well informed "apartisans" (who are always considering changing their vote) constitute an important part of the electorate [30].

In addition to the relationship between Wikipedia page views and electoral outcomes, any theory of online information seeking and politics also needs to take into account the potential influence of the news media, which has traditionally been the venue where people seek politically relevant information [31]. Information patterns within the news media may themselves correlate with electoral outcomes to a significant extent, as media outlets will often seek to make their coverage more or less proportional to the importance of different political parties. However, coverage of older parties is perhaps likely to be more widespread, especially if they are incumbent in the current government. News media coverage may also have an impact on Wikipedia page views, in both a positive and negative sense: being mentioned in the news media might stimulate people to find out more about a candidate online; but it might also fulfil people's information needs, and hence make them less likely to seek information from alternative sources.

This theory of how online information seeking relates to eventual outcomes suggests that a model which uses Wikipedia page views to predict electoral outcomes should take into account several factors. First, we would expect new parties, already established parties and incumbent parties to experience different page view dynamics, with newer parties receiving a disproportionately large amount of page views compared to their final number of votes (a result already observed in socially generated predictions based on Twitter data [19]), whilst more established parties experience the reverse effect. Second, parties attracting lots of swing voters may also do disproportionately well in page view statistics, whilst parties which are losing votes may again do disproportionately badly. Finally, coverage of the political party in the mainstream news media may well correlate with voting behaviour itself, as well as serving to "replace" any effect of Wikipedia page views (hence parties which are well covered by the news media are may be badly covered by Wikipedia).

## 2. Data

Our aim in this paper is to test the extent to which models can be developed using Wikipedia data to predict electoral outcomes (both in terms of turnout and especially in terms of results for individual parties), based on these simple theoretically informed corrections. In order to do this, we built a dataset centred on the two most recent European Parliament elections (in 2009 and 2014). The European Parliament elections were chosen because they would allow us to build a relatively large sample of political parties, all competing at the same time under broadly the same electoral system. However, it should be noted that this focus is somewhat of a limitation as European elections are typically perceived as secondary elections in most EU member states, subordinate to the national electoral contest: expansion to national elections would be a useful next step.

We collected two types of data on these elections. First, page view statistics for the general Wikipedia page on the election were harvested in 14 different language editions of Wikipedia, each one representing one of the countries which went to the polls on election day (for example, the relevant page in the English version of Wikipedia for the 2009 EU Parliament Elections is: http://en.wikipedia.org/wiki/European_Parliament_election,_2009).

These 14 language editions were chosen based on the following criteria: (a) they are primarily spoken in one country, (b) the country that the language is spoken in has been an EU member in the last two rounds of elections, and (c) the corresponding Wikipedia pages have been existing prior to the election date in that language edition. They were also chosen because they have relatively strong user bases in Wikipedia, which provides a good basis for comparison.

Second, we created a dataset of political parties which competed in either or both of these elections in the five largest Western European countries: the UK, France, Germany, Spain and Italy. Parties which competed in both are represented twice, with one observation for each election. The full list of parties is available in the Additional File 1. As we have highlighted above, we expect the perceived viability of political parties to be a necessary condition for generating significant levels of Wikipedia traffic. Hence we chose to focus only on those parties which secured more than 5% of the vote in either year (even if a party scored 5% in 2009 and 4% in 2014, it is still included for both elections). European Parliament elections (like any election) typically feature a small number of parties which absorb the vast majority of votes, and a long tail of more minor parties which achieve little or no electoral success [22]. Having such a threshold is therefore also a necessary simplification, as it removes a long tail of very minor parties who otherwise would have constituted the majority of the observations. In total the dataset contains 59 observations.

For each party, we recorded several variables of interest. First, we recorded the amount of views the page of the political party in question in the corresponding Wikipedia language edition received in the week before the election. There was some ambiguity about which Wikipedia page to use in some cases, especially where more than one party presented itself in the form of a coalition, and hence more than one page could be valid. In these cases, we always used the Wikipedia page which had the highest number of views. A full list of the Wikipedia pages used is available in Additional File 1. We then also recorded the percentage of the vote achieved by the party, and the difference between that percentage and the previous year (which captures whether voters swung towards or away from the party). Results of the 2009 and 2014 elections were taken from http://www.results-elections2014.eu/. Results of the 2004 elections, which were needed to calculate the change between 2004 and 2009, were taken from http://www.nsd.uib.no/.

We also recorded several variables which allow us to apply our theoretically informed corrections. First, we recorded whether a party is "new" or not. Newness is a somewhat ambiguous category, as many of the parties which surged to prominence in the 2014 elections had existed for a long time as relatively minor political forces, for example, UK Independence Party (UKIP), and many apparently new parties are rebrandings of existing ones. Our aim with this variable was to capture the extent to which the majority of the electorate is likely to recognise this party already. Hence parties were listed as new not only if they did not compete at the previous election, but also if they either had a different name to that which they competed under in the previous election, or scored less than 5% at the previous election. We made two exceptions to this rule: the major centre left and centre right parties in Italy both changed their names before the 2009 election; however, we did not record them as "new" parties because they were the major political forces in Italy at the time. Incumbency was a simpler variable to measure: it simply records whether the party was an incumbent in the national government at the time of the election, which would again likely change the extent to which they are visible to the electorate.

Finally, we recorded the amount of times the party was mentioned in the print media in the week immediately prior to the election. These numbers were calculated by conducting a search for the party on LexisNexis' news media dataset (http://www.lexisnexis.co.uk/) in the largest official language of the country in question. This dataset is a large archive of material produced by print newspapers all over the world, which can act as a useful proxy of media attention to an individual party. The number of search results returned was used as the number of media "mentions" recorded (with one mention being one news article which contained at least one reference to the party in question). This variable is of course not a perfect measure. In particular, in many cases the exact search term to use in LexisNexis was not self-evident. For example, the Christlich Demokratische Union in Germany is often referred to as the CDU. In cases where more than one search term was potentially valid, we conducted the search with all possible terms, and used the highest number which resulted. We did however avoid terms which are also commonly used words, for example "greens" as a shorthand for the Green Party. A full list of search terms used is available in the Additional File 1.

## 3. Results

Our results section is divided into three parts. First, we look at the relationship between Wikipedia traffic patterns around election time and overall electoral turnout. Then, we develop a model to try and predict absolute vote share outcomes for different political parties. Finally, we look at whether such a model can be developed for swing voters, measured in terms of relative change in vote share.

We will begin by looking at the relationship between overall levels of traffic to the general political articles in different language editions of Wikipedia and electoral turnout. These articles offer general information about the election, such as the date on which it will be held. As we highlight in our theory section we expect that more people looking at this page indicates that higher levels of people are interested in voting in the election, and are seeking to inform themselves about the practicalities of voting.

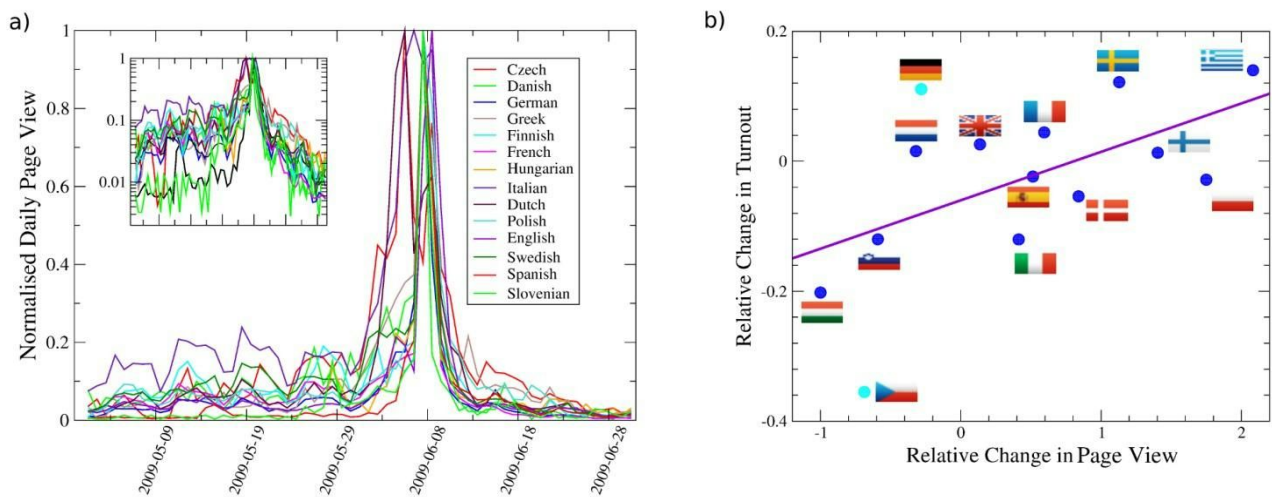

**Figure 1. EU elections and Wikipedia page views.** a) Page views of the general Wikipedia article on the European Parliament Elections over time and b) compared to change in turnout. The two outliers, Germany and Czech Republic are excluded from correlation analysis.

Figure 1(a) shows the overall pattern of page views of the main Wikipedia article in the EU parliament elections around the election time in 2009 in 14 different language editions of Wikipedia. Despite all the differences in the political settings of the associated countries, some common patterns are evident from this diagram: a gradual increase can be observed, starting about a month before the election date, followed by a large jump a few days before the election and finally a decay after the results are announced. It seems that the vast majority of information seeking about elections takes place in the few days before the election itself. These data further justify a focus on page views in the week just before the election. From the inset of Figure 1(a), which shows the same data in logarithmic scale, we observe that the absolute value of the decay rate in the post-event attention is larger than the one of the pre-election build up phase (similar to observations in [23]). In other words, interest in the event fades very quickly. The peaks of attention fall on different days: we attribute this to the fact that the EU election falls on different days (and sometimes spans multiple days) in different countries.

If our theory of the reasons for online information seeking holds true, then the volume of attention in the build-up phase before the election may be an indicator of the general level of interest in that election (in particular, whether people are considering a vote at all), and therefore a predictor for the overall turnout in each country. Of course, different language editions of Wikipedia have very different sized user bases, meaning that the absolute level of attention to the general Wikipedia article is not likely to be of much use as a predictor. Hence we look instead at the relative change in page views to this general Wikipedia article between the 2009 and 2014 European elections, and compare this to the relative change in turnout between those two elections. Figure 1(b) plots the correlation between these two values. Once we remove two outlying points, the correlation of $R$= 0.72 (Adjusted $R^2$ 0.47, p-value = 0.004) is surprisingly high considering both the limitations of the data and the simplicity of the model. This shows some good initial support for our theory that general levels of interest in a political event are proportional to general levels of readership on Wikipedia. However, we do not have a good explanation for the different dynamics which may have produced the two outliers.

We now move on to our main focus, which is predicting the performance of individual parties in the two elections and five countries as described above. We focus again on page views in the week before the election, this time to the Wikipedia page of the individual party in question in the language edition relevant to that country. Again, as language editions of Wikipedia have different volumes of traffic, we conduct another normalisation, but this time to the sum of page views of all parties competing in the same country at the same time. This can be thought of, in other words, as a party's "share" of Wikipedia traffic for that election.

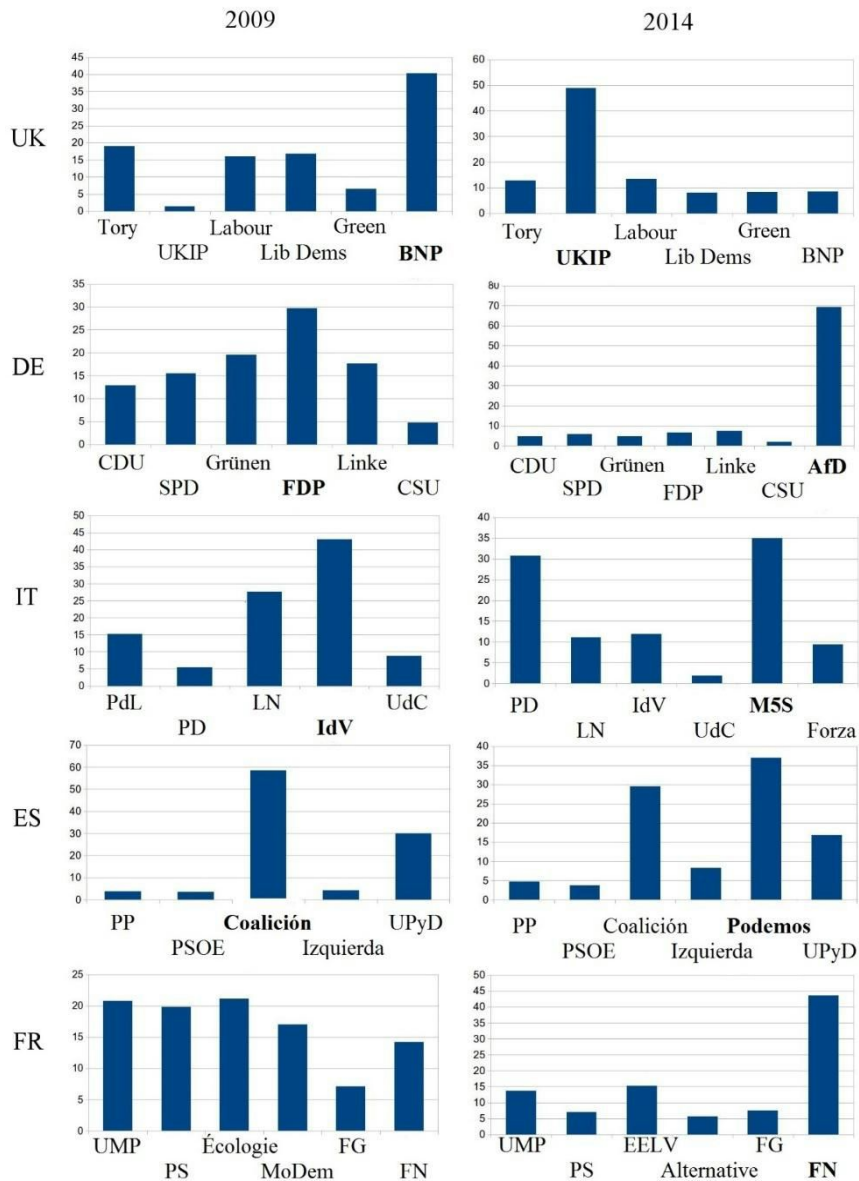

**Figure 2. Parties' Wikipedia page views.** Relative share of Wikipedia traffic for major parties in five European countries the week before the 2009 and 2014 European parliament elections (for the party names, see the list of abbreviations below).

As we describe above, theoretically we do not expect much of a correlation between absolute levels of Wikipedia traffic and absolute levels of vote share (and indeed, in practice, we observe $R$=0.05). We explore the reasons for this in Figure 2, which shows the relative levels of traffic each party achieved, separated for each country/election. We observe that the "winner" in terms of Wikipedia vote share is, of course, not the winner in terms of overall vote share, but is instead often an "anti-establishment" party who ended up doing surprisingly well. This is the case of, for example, the Front Nationale (FN) in France in 2014, the British National Party (BNP) in the UK in 2009, Podemos in Spain in 2014, etc. This fits in with some of the theory we described above: these parties were relatively minor political forces at the time, and some were completely new, but they attracted a lot of swing voters in those elections. Hence they were doubly likely to be favoured in page view statistics: lots of voters were considering voting for them, and many of these voters would have had little prior information about these parties.

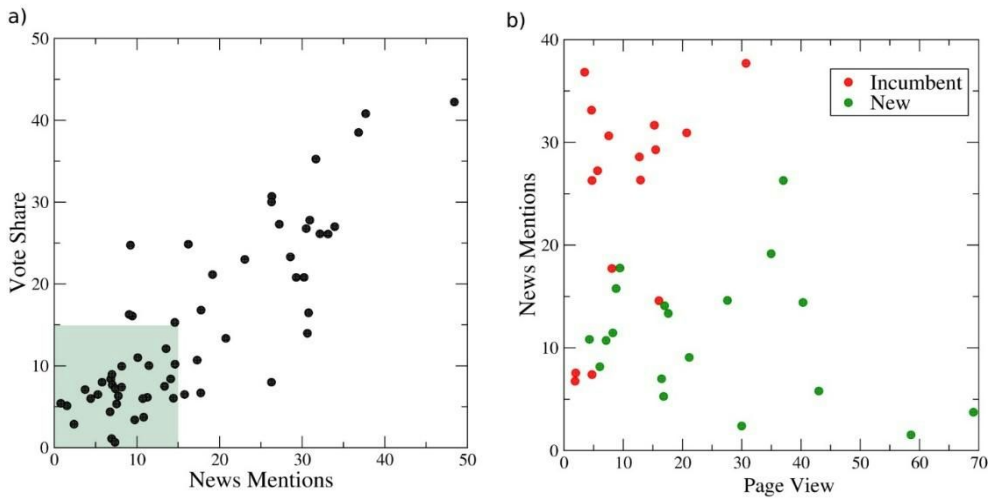

**Figure 3. Attention and Media affect.** News media mentions compared to (a) Wikipedia page views and (b) absolute level of vote share.

As we highlight in our theoretical discussion, we expect several corrections to be necessary to Wikipedia view data in order for it to act as an effective predictor. We will examine first of all the news media. Figure 3(a) shows the correlation between news mention counts of political parties in the week before the election and the overall vote share. The correlation is actually very high ($R$=0.84, Adjusted $R^2$=0.71, p-value < 0.00001). As we described above, news outlets tend to calibrate their coverage to the perceived importance of political parties, hence this correlation is not too surprising. However, it is notable that it is also much stronger for the larger parties, whilst only moderate for the smaller parties shown in the shaded area of the graph ($R$=0.52, Adjusted $R^2$=0.25, p-value = 0.0005).

We have observed that Wikipedia predicts the successful emergence of new parties very well and news media is a good indicator for the more established parties, we can further deepen this observation by comparing Wikipedia attention to news mentions. We suggested above that the news media might drive traffic to Wikipedia (which would make Wikipedia page views essentially epiphenomenal), or that it might also "replace" Wikipedia (with more news mentions leading to less views. However, as we see in figure 3(b), neither of these correlations can be observed. In figure 3(b), parties are colour coded based on their political history: green for new parties and red for incumbent parties (with parties which were neither new or incumbent removed). We can see that there are two distinct clusters on this graph: incumbent parties which are over-represented in the media and under-represented in Wikipedia, and vice versa for new parties. This would suggest that news media mentions should be corrected to account for incumbency, whilst Wikipedia mentions are corrected to account for the newness of a party.

With these observations in hand, we will now move on to a more systematic investigation of the usefulness of Wikipedia page view data in predicting electoral outcomes. We divide this investigation into two parts. First, we look at the use of page view data in predicting absolute vote share outcomes for individual parties. Second, we look at the use of this type of data in predicting changes in vote share outcomes when compared to previous elections.

We will begin with our observations about absolute vote share outcomes. As we highlighted above, we theorise that increases in Wikipedia page view counts may indicate increasing numbers of votes for a party, as voters who are considering a vote for that party may be informing themselves about its policies. However, we also expect a number of corrections to be necessary. We test this contention with four analytical models reported in table 1 which seek to estimate the vote share (measured from 0-100) for each party in the dataset. These models are linear OLS regressions looking at the relationship between voting outcomes and predictors related to information seeking behaviour, and were estimated in R. The first model, model 1.0, is a baseline model, which contains all theoretically relevant parameters except those relating to Wikipedia. In particular, it contains parameters for the relative share in the news items (measured from 0 – 100), a dichotomous indicator for whether the party is a new party or an incumbent party, and an interaction effect between the news count and the incumbency indicator. This model serves as a basis for comparison for model 1.1, which is a full model containing an additional two Wikipedia relevant parameters: a main effect for the relative share in Wikipedia page views (again measured from 0 - 100), and an interaction effect between Wikipedia views and status as a new party. Models 1.2 and 1.3 duplicate models 1.0 and 1.1, but looking only at the subset of parties with less than 15% of the vote. These parties were highlighted in the bottom left hand corner of figure 3(a) as observations where the "news only" model seemed to perform poorly, and also observations where, following figure 2, Wikipedia might offer useful information.

| † p < 0.01<br>* p < 0.05<br>** p < 0.01<br>*** p < 0.001 | Model 1.0:<br>Absolute vote<br>share (baseline) | | Model 1.1:<br>Absolute vote<br>share (full model) | | Model 1.2:<br>Small parties only<br>(baseline) | | Model 1.3:<br>Small parties only<br>(full model) | |
|---|---|---|---|---|---|---|---|---|
| | $β$ | SE | $β$ | SE | $β$ | SE | $β$ | SE |
| Intercept | 3.66* | 1.81 | 1.96 | 2.13 | 4.75*** | 1.09 | 2.58 | 1.59 |
| News | 0.66*** | 0.09 | 0.65*** | 0.09 | 0.24* | 0.09 | 0.25** | 0.09 |
| New Party | -2.00 | 1.72 | -1.15 | 2.72 | 0.00 | 0.94 | 1.79 | 1.66 |
| Incumbency | -6.24 | 3.85 | -4.91 | 3.94 | -1.55 | 2.37 | 0.18 | 2.50 |
| News x Incumbency | 0.35* | 0.16 | 0.31† | 0.16 | 0.10 | 0.16 | 0.05 | 0.16 |
| Wikipedia | | | 0.12 | 0.08 | | | 0.16† | 0.09 |
| New Party x Wikipedia | | | -0.09 | 0.11 | | | -0.15 | 0.09 |
| $R^2$ | 0.75 | | 0.76 | | 0.33 | | 0.40 | |
| Adjusted $R^2$ | 0.73 | | 0.73 | | 0.24 | | 0.27 | |
| N | 59 | | 59 | | 35 | | 35 | |

**Table 1:** Predicting absolute vote share outcomes

Models 1.0 and 1.1 provide little evidence for the idea that Wikipedia page view data is of use in predicting absolute vote share outcomes. The overall $R^2$ of these models is comparatively high. However, this is driven largely by the news coefficient. When compared to the baseline model, model 1.1 increases $R^2$ by just 0.01, which translates to a fractional increase in adjusted $R^2$ (of 0.002). Models 1.2 and 1.3 provide somewhat more evidence of Wikipedia's usefulness, though still in relatively modest terms. Both models do a relatively poorer job at predicting the results of small parties, when compared to large ones. However, the $R^2$ of model 1.3 is 0.07 greater than that of 1.2, which translates into an improvement in adjusted $R^2$ of 0.03. Hence Wikipedia data appears to be of somewhat more use for smaller parties. This is consistent with our observations above about Wikipedia providing a pointer to the successful emergence of smaller parties.

It is also worth discussing the coefficients of individual indicators. As we would expect based on figure 3(a), the news coefficient is positive, statistically significant in all models, and of a considerable size: for example, gaining 50% of the news mentions would, on average, correlate with an increase in vote share of around 33 percentage points. This relation is increased for incumbent parties, as shown by the interaction term between incumbency and news mentions (though the interaction term is only statistically significant in the first two models). The term for Wikipedia also points in the expected direction in all models: more Wikipedia page views do seem to correlate with more votes, though the size of the coefficient is much smaller (50% of the Wikipedia "page view share" would correlate with around 6 percentage points more votes). As we also expected, the new party interaction reduces the effect of the Wikipedia term considerably. Hence it does indeed seem to be the case that Wikipedia page views overstate the potential electoral impact of very new parties. We attribute this to the fact that higher levels of information seeking are rational for newer parties which are less well known, as discussed in the theoretical section. By contrast, we do not find any support for the "viability" thesis (that is, the idea that new parties will get attract a disproportionately small amount of information seeking as people do not perceive them as a viable option). However it should be noted that the term for Wikipedia is only on the borderlines of statistical significance in model 1.3 (p=0.08), and is not significant in model 1.1, whilst the new party interaction term is not significant.

| † p < 0.01<br>* p < 0.05<br>** p < 0.01<br>*** p < 0.001 | Model 2.0:<br>Change in vote share (baseline) | | Model 2.1:<br>Change in vote share (full model) | | Model 2.2:<br>Small parties only (baseline) | | Model 2.3:<br>Small parties only (full model) | |
|---|---|---|---|---|---|---|---|---|
| | β | SE | β | SE | β | SE | β | SE |
| Intercept | -0.70 | 2.45 | -6.45* | 2.51 | -2.43 | 1.54 | -5.71* | 2.14 |
| News | -0.02 | 0.12 | -0.03 | 0.10 | 0.06 | 0.13 | 0.13 | 0.12 |
| New Party | 3.35 | 2.33 | 5.29 | 3.19 | 3.91** | 1.33 | 4.10† | 2.23 |
| Incumbency | -3.27 | 5.22 | 1.21 | 4.64 | 1.36 | 3.34 | 4.05 | 3.36 |
| News x Incumbency | 0.15 | 0.22 | 0.03 | 0.19 | -0.13 | 0.22 | -0.26 | 0.21 |
| Wikipedia | | | 0.40*** | 0.10 | | | 0.21† | 0.12 |
| New Party x Wikipedia | | | -0.25† | 0.13 | | | -0.12 | 0.12 |
| $R^2$ | 0.05 | | 0.32 | | 0.25 | | 0.40 | |
| Adjusted $R^2$ | -0.02 | | 0.24 | | 0.15 | | 0.27 | |
| N | 59 | | 59 | | 35 | | 35 | |

**Table 2:** Predicting relative vote share outcomes

As we highlighted above, we also expected Wikipedia to offer predictive power in the case of swing voters rather than in the case of absolute vote share. Hence we will now move on to a second set of analytical models which try and address this question, which are presented in table 2. Instead of trying to predict absolute amounts of vote share, these models try and predict change, that is the amount by which a party's vote share increases or decreases compared to the last election (with new parties considered to have been at 0 votes in the last election). This variable is again measured from 0-100. Apart from this new dependent variable, models 2.0-2.3 are identical to models 1.0-1.3.

These models provide much better evidence for the potential usefulness of Wikipedia page views. The baseline model (2.0) performs very poorly indeed, with a negative adjusted $R^2$. However, the full model, 2.1, performs much better, with an adjusted $R^2$ of 0.24. Hence when it comes to predicting relative change in vote share, Wikipedia seems to provide a reasonable amount of information (though it should be stated that the overall $R^2$ is still comparatively low). The coefficient for the Wikipedia term is also statistically significant, with parties which hold 50% of the Wikipedia share for a given country picking up, on average, a 20 percentage point increase in votes. The new party term reduces this effect considerably, providing further evidence that Wikipedia page overstates new parties.

Models 2.2 and 2.3 are again duplicates of 1.2 and 1.3 (that is, only looking at small parties) with this new dependent variable. The results are similar to those found in models 2.0 and 2.1. The term for new parties becomes statistically significant and leads to some improvement in adjusted $R^2$ in the baseline model, when compared to model 2.0. However, this is an artifice of the data: as a new party cannot achieve a negative change in vote share when compared to the previous election, new parties are of course more likely to have positive swings than more established parties. In model 2.2 we can see the continued impact of the Wikipedia variables, which serve to increase the adjusted $R^2$ from 0.15 in model 2.3 to 0.27 in model 2.4. However, this model for small parties is not a major improvement on the model which looks at all parties. In other words, when it comes to change in vote share, it seems that Wikipedia offers just as much information for large parties as it does for small parties.

## 4. Discussion and Conclusions

In this paper we have sought to develop theoretically informed methods for election prediction based on information seeking behaviour on Wikipedia, responding to existing critiques of predictions generated from new sources of big data. We applied this model to a variety of different European countries in the context of two different European elections. We have produced three main findings. First, we have shown that the relative change in the number of page views to the general Wikipedia page on the election can offer a good estimation of the relative change in turnout for that election at the country level. This supports the idea that increases in online information seeking at election time are driven by voters who are considering voting in the election. Second, we have shown that a theoretically informed model based on the Wikipedia page views, news media mentions, and basic information about the political party in question can offer a good prediction of the overall vote share of the party in question. However, we also found that the majority of this predictive power came from the news media variable, rather than Wikipedia itself, though Wikipedia views appear to offer a correction for small parties. Third, we presented a model for predicting change in vote share (i.e. voters swinging towards and away from a party). Though overall $R^2$ for such a model was comparatively low, we showed that Wikipedia page view data provided for an important increase in predictive power in this context.

Based on this information, we can also enhance our theory of the sources of Wikipedia traffic before election time (and hence potentially improve future forecasting models). We have shown good evidence that newer parties which attracted a lot of swing voters received disproportionately high levels of Wikipedia traffic, offering further support to the idea that voters are cognitive misers who are more likely to seek information on new political parties and when they are actually changing a vote. By contrast, there was no evidence of a "media effect": there was little correlation between news media mentions and overall Wikipedia traffic patterns. Indeed, the news media and Wikipedia appeared biased towards different things: with news favouring incumbent parties, whilst Wikipedia favoured new ones. As we continue to understand more about what drives these traffic patterns, predictions based on the data have the potential to increase in accuracy.

It is worth concluding by reflecting on the limitations of the work. We focussed on EU level elections: national elections may experience different dynamics, and would also be worth studying. Our measure of swing voters (change in vote share when compared to the previous year) is also imperfect, as voters may of course swing both towards and away from a party in an election. Furthermore, we do not know much about the content of the news article used: whether it was short or long, positive or negative. Such data would undoubtedly improve our understanding of the informational landscape surrounding parties at election time. Likewise, we do not know to what extent Wikipedia page views really are driven by voters, as opposed to other political actors such as journalists. Perhaps most importantly, our theories and conclusions concern micro level behaviour, but our data is measured at an aggregate level, which may mask micro level effects. Future work could usefully address these concerns.

## Electronic Supplementary Material

Additional file 1, .csv, **Party List**: A table containing countries, name of the parties (English and local), election dates, party abbreviations, election vote share, change in the vote share from the previous election, number of news mentions, and the link to the Wikipedia page.

**List of abbreviations**

AfD: Alternative für Deutschland

BNP: British National Party

Grünen: Bündnis 90 / Die Grünen

CDU: Christlich Demokratische Union

CSU: Christlich-Soziale Union

Coalición: Coalición por Europa

Tory: Conservative Party

Linke: Die Linke

Écologie: Europe Écologie

EELV: Europe Écologie Les Verts

Forza: Forza Italia

FDP: Freie Demokratische Partei

FG: Front de gauche

FN: Front National

Greens: Green Party

PdL: Il Popolo della Libertà

IdV: Italia dei Valori

Izquierda: La Izquierda

Labour: Labour Party

LN: Lega Nord

Lib Dems: Liberal Democrats

MoDem: Mouvement Démocrate

M5S: Movimento 5 Stelle

PS: Parti Socialiste

PP: Partido Popular

PSOE: Partido Socialista Obrero Español

PD: Partito Democratico

Podemos: Podemos

SPD: Sozialdemokratische Partei

UMP: Union pour un mouvement populaire

UPyD: Unión Progreso y Democracia

UdC: Unione di Centro

UKIP: United Kingdom Independence Party

**Competing interests**
The authors declare that they have no competing interests.

**Authors' contributions**
Both authors participated in the design of the study, performed the statistical analysis, and drafted the manuscript. Both authors read and approved the final manuscript.

**Authors' information**
TY is a Research Fellow in Computational Social Science at the Oxford Internet Institute of University of Oxford. He is a physicist by training. JB is a Research Fellow at the Oxford Internet Institute of University of Oxford. He is a political scientist specialising in computational and 'big data' approaches to the social sciences.

**Acknowledgements**
This publication arises from research funded by the John Fell Oxford University Press (OUP) Research Fund; Grant Number: CZD06990.